\documentclass[a4paper,12pt,oneside]{article}

\usepackage{amsmath}
\usepackage{amssymb}
\usepackage{amsthm}
\usepackage{t1enc}
\usepackage{verbatim} 
\usepackage{graphicx}
\usepackage[dvips]{epsfig}
\usepackage{psfig}
\usepackage{url}
\usepackage{booktabs}
\usepackage{float}
\usepackage{algorithmic}
\usepackage[usenames, dvipsnames]{color}
\definecolor{mygray}{gray}{0.6}

\theoremstyle{plain} 

\theoremstyle{definition} %

\theoremstyle{remark} %

\begin{document}
\begin{center}
	\LARGE A parallel space-time multigrid method for the eddy-current equation\footnote{The research was funded by the Austrian Science Fund (FWF) under the grant W1214-N15, project DK8.}\\~\\
	\normalsize Martin Neum\"{u}ller\footnote{Herzogsdorf, Austria, E-mail: martin@neumuller.at}
	~~~~~~~ Martin Schwalsberger\footnote{Doctoral Program "Computational Mathematics", Johannes Kepler University Linz, Altenberger Stra\ss{}e 69, A-4040 Linz, Austria, 
		E-mail: martin.schwalsberger@dk-compmath.jku.at}\\~\\
	October 30, 2019
\end{center}

\begin{abstract}
	We expand the applicabilities and capabilities of an already existing space-time parallel method based on a block Jacobi smoother. First we formulate a more detailed criterion for spatial coarsening, which enables the method to deal with unstructured meshes and varying material parameters. Further we investigate the application to the eddy-current equation, where the non-trivial kernel of the curl operator causes severe problems. This is remedied with a new nodal auxiliary space correction. We proceed to identify convergence rates by local Fourier analysis and numerical experiments. Finally, we present a numerical experiment which demonstrates its excellent scaling properties.
\end{abstract}



\section{Introduction}

Around 2004 processor speed ceased to increase due to physical limits. Therefore we have experienced exponential growth in the number of processing units instead. This manifests in modern processing clusters featuring 1000 to more than ten million CPUs as of 2019\footnote{\url{https://www.top500.org/}, accessed 2019}. Meanwhile applications that are parallel in space frequently fail to efficiently utilize more than a few thousand CPUs (depending on problem size).  To utilize this processing power massively parallel algorithms are required. For time dependent simulations the time dimension offers potential for further parallelization, however the classical time-stepping concept is fundamentally sequential. Primarily for parabolic evolution equations time-parallel methods have been developed to break this "curse of sequentiality". After the first concept for ordinary differential equations more than 55 years ago \cite{MR0176617}, a whole family of time-parallel methods of all kinds has blossomed, see \cite{MR3676210} for a historical overview. \\

Despite still being sequential in time the method \cite{MR806780} by Hackbusch laid some groundwork for time-parallel multigrid schemes and defied the notion of exactly solving the problem forward in time. Lubich and Ostermann took a different approach by presenting a waveform relaxation type space-time multigrid method \cite{MR894124}. An in depth discussion of waveform relaxation methods 
can be found in the monograph \cite{MR1231343}. A further variant described as space-time concurrent multigrid waveform relaxation with cyclic reduction can be found in \cite{MR1326811}.\\

A contemporary method similar to ours is multigrid reduction in time \cite{MR3499068} which has been actively expanded by a team centered around the Lawrence Livermore National Laboratory \cite{MR3499068,MR3716570,MR3716560,7741520,MR3830280,8586435}. Further results can be found on the XBraid project website\footnote{\url{github.com/XBraid/xbraid/wiki/Project-Publications}}. Another very active group is the J\"ulich Forschungszentrum with diverse publications about time-parallel methods \cite{6468522,Speck:151996,Speck:171823,Speck:155430,Steiner:185897,MR3345221,Ruprecht:202567,Speck:202569,MR3504550,MR3411259,MR3537015,MR3612601,MR3730032,MR3830279,MR3890987}. \\

A method, which can be described as the precursor to the next method, is space-time multigrid with point-wise relaxation \cite{MR1335894}. The method used in this paper  \cite{MR3521549} replaces the point-wise relaxation with block relaxation. Thus it can be described as  space-time multigrid with block relaxation. 
Recent publications demonstrated that this method can also be applied to stochastic differential equations \cite{MR3815550}, and in combination with isogeometric analysis methods \cite{MR3819379}
. This motivated us to investigate the application of our method to the eddy-current equation. Furthermore we generalize concepts to expand our scope towards more complex problems.\\

The eddy-current equations 
are used to study low frequency electromagnetic applications, like electric motors/generators, induction furnaces, non-destructive testing, and many more. In practice, time-harmonic approaches are often preferred over transient models due to lower computational costs. However time-harmonic approaches struggle with nonlinear materials (saturation effects) and moving geometry. As a stepping stone to these more complex problems, we only consider the linear case with constant geometry in this paper.\\


The remainder of this paper is organized as follows. In Section 2  the Space-time-parallel method is introduced and the concept of a coarsening rule is presented and expanded.
Section 3 applies the method to the eddy-current equation and pinpoints the core problem of the naive approach. We continue to determine a remedy to this problem - namely a "nodal auxiliary space correction"-, which can be used to define a hybrid smoother.
In Section 4, we perform a local Fourier analysis of the introduced hybrid smoother for the 2D curl-curl equation.
In Section 5, we mimic the local Fourier analysis with numerical experiments for the 3D eddy-current equation. We further investigate the validity of a new generalized coarsening rule.
Weak scaling results of the improved method are given in Section 6.
Conclusions are drawn in Section 7 and we discuss possible implications of our findings for related models and methods.

\section{The Method}
\subsection{The Space-Time Parallel Method}

For simplicity we reiterate the space-time parallel multigrid method presented in \cite{MR3521549} and \cite{MasterSchwalsberger} only for the case of the implicit Euler scheme with uniform step width $\tau=T/m$. It should be noted, that we are not concerned with the approximation quality, but we only want to solve a given discretized problem. After the usual discretization steps for a PDE in space and time (see Chapter \ref{Eddy-Currents} for an example) we need to solve an equation of the form

\begin{equation}\label{timeStepping}
	-M_h \underline{u}_h^k + (M_h+\tau K_h)\underline{u}_h^{k+1} = \tau \underline{f}_h^{k+1},~~~~k=0,...,m-1, ~~~~ M_h\underline{u}_h^0=\underline{u}_{0,h}.
\end{equation}

By defining $A_{\tau,h}:=M_h+\tau K_h$ we can rewrite the time-stepping scheme as a huge linear system
\begin{equation*}
\begin{pmatrix}
A_{\tau,h} &  &  &  \\ 
-M_h & A_{\tau,h} &  &  \\ 
& \ddots & \ddots &  \\ 
&  & -M_h & A_{\tau,h} \\ 
\end{pmatrix} 
\begin{pmatrix}
u_h^1\\ 
u_h^2\\
\colon\\
u_h^m
\end{pmatrix} 
=
\begin{pmatrix}
\tau f_h^1 + u_{0,h}\\ 
\tau f_h^2\\
\colon\\
\tau f_h^m
\end{pmatrix}.
\end{equation*}

This system can be compactly written as
\begin{equation*}
L_{\tau,h}\underline{x}=\underline{f}.
\end{equation*}

The standard smoother in this method is a block-Jacobi preconditioned Richardson scheme with $\omega=0.5$
\begin{equation}\label{StandardSmoother}
	\underline{x}^{k+1}= \underline{x}^k+\omega\underbrace{(I\otimes A_{\tau,h}^{-1})}_{=\text{diag}(A_{\tau,h}^{-1},...,A_{\tau,h}^{-1})}(f-L_{\tau,h}\underline{x}^k).
\end{equation}
 
The overall method consists of pre-smoothing, a recursive coarse time-grid correction and post-smoothing. This V-cycle over the time meshes is guarantied to geometrically converge, as long as the application of $A_{\tau,h}^{-1}$ is sufficiently well approximated and $M_h^{-1}K_h$ is positive semidefinite with real eigenvalues.\\

The presented smoother is simultaneously parallel in space and time. Because all operations are a Kronecker product of a time operation and a spatial operation, all spatial operations can be executed with the usual space-parallel approaches. On top of this also the time mesh can be subdivided among groups of processors. Only the application of $L_{\tau,h}$ requires forward communication for neighboring time nodes.

\subsection{Spatial Coarsening}
For performance reasons it is desirable to use not only a coarser time-mesh for the coarse grid correction, but also a coarser spatial mesh. For example without spatial coarsening the coarse grid correction is responsible for approximately half the computational effort and is the less parallel part of the algorithm. In 3D the use of a coarser spatial mesh approximately reduces the effort for the coarse grid correction by a factor of $1/8$ almost doubling the overall speed of the method.\\

However various inhibitors can prevent proper convergence rates for spatial coarsening. The "degree of anisotropy" in the discrete problem (denoted by $\lambda$, coined in \cite{MR1335894}) is very significant for the behavior of the method. Assuming properly working spatial solvers the degree of anisotropy is indeed the only potential inhibitor for the heat equation.
In fact if $\lambda$ is above a certain threshold and no other inhibitor is active, then spatial coarsening has no negative effect on the convergence rate. Therefore we use spatial coarsening only if the \textit{coarsening rule} is fulfilled
\begin{equation*}
	\lambda:= \frac{\tau}{h^2} \geq \lambda_{crit},
\end{equation*}
with $\tau$ being the time step size and $h$ the spatial mesh size.
The threshold $\lambda_{crit}$ depends on the used time-stepping scheme, the spatial discretization method, the smoothing steps, the dimension and the underlying differential equation. This rule needs to be generalized for practical applications to generic parabolic PDEs with a generic differential operator $D_x$
\begin{equation*}
	\alpha \frac{\partial u}{\partial t} + D_x^* \beta D_x u = f.
\end{equation*}
By a variable substitution argument we can account for the parameters $\alpha, \beta$ and by defining a local degree of anisotropy we can account for variation of the values over the spatial domain. The final coarsening rule is defined as
\begin{equation} \label{CoarseningRule}
	\lambda(\mathcal{T}):= \underset{x\in \mathcal{T}}{\min} \frac{\beta(x)\tau}{\alpha(x)h_\mathcal{T}^2} \geq \lambda_{crit}~~~~~~\forall \mathcal{T}\in T_h(\Omega).
\end{equation}
So for each element of the spatial mesh $T_h(\Omega)$ this condition needs to be satisfied for spatial coarsening. However in practice with piecewise constant parameters and other assumptions simplifications for checking this rule can be made as explained in \cite{MasterSchwalsberger}. To measure $h_\mathcal{T}$ we decide to use the length of the longest edge of the tetrahedron, whereas equivalent choices would be reflected in the value of $\lambda_{crit}$.\\

The determination of $\lambda_{crit}$ in each case can be done either by a local Fourier analysis (LFA) or by numerical experiments. Both of these approaches are demonstrated in this paper.

\section{Eddy-Currents}\label{Eddy-Currents}
After the successful application of this scheme to the heat equation \cite{MR3521549},
we attempt to solve the eddy-current equation
\begin{equation}\label{EddyCurrent}
	\sigma(x)\frac{\partial \underline{u}}{\partial t}+\text{curl}~\mu^{-1}(x)\text{curl}~\underline{u}=\underline{f}, ~~~~~~\text{in}~ \Omega\times (0,T).
\end{equation}
under the regularization assumption $\sigma(x) \geq 1[\frac{S}{m}]$ to avoid dealing with differential algebraic equations for now. 
We also always use homogeneous Dirichlet ($u\times n=0$,~~on $\partial\Omega\times (0,T)$) or Neumann ($\mu^{-1} \text{curl} (u) \times n = 0$,~~on $\partial\Omega\times (0,T)$) boundary conditions and a homogeneous initial condition ($u= E_0$,~~ for $t=0,~x\in \Omega$). Applying an implicit Euler scheme and lowest order Nedelec elements \cite{MR2059447} to the semi-variational formulation
\begin{equation*}
	\frac{\partial}{\partial t}\int_\Omega \sigma \underline{u}(t)\cdot \underline{v} dx +\int_\Omega\mu^{-1}\text{curl}~\underline{u}(t)\cdot \text{curl}~\underline{u} dx = \int_\Omega \underline{f}(t)\cdot \underline{v} dx ~~~~ \forall \underline{v} \in H(\Omega,\text{curl})
\end{equation*}
yields the discrete time-stepping scheme (\ref{timeStepping}), where
the occuring matrices are defined by the basis $(\phi_i)$ of the Nedelec finite elements
\begin{equation*}
	(M_h)_{i,j} := \int_\Omega \sigma \phi_i \cdot \phi_j dx,~~~~	(K_h)_{i,j} :=  \int_\Omega\mu^{-1}\text{curl}~\phi_i\cdot \text{curl}~\phi_j dx 
\end{equation*}

\subsection{Troubles with Spatial Coarsening}

As $M_h^{-1}K_h$ is positive semidefinite and has real eigenvalues, it is not surprising, that the pure time multigrid method without spatial coarsening works reasonably well with convergence rates mostly below $0.5$. For the idealized case of  an exact solver for $A_{\tau,h}$ this rate is proven by inspecting the test equation (\ref{TestEquation}) as in \cite[ch.4.2]{NeumDiss}. However for the method as it is we come to the conclusion: \textit{A coarse grid correction with spatial coarsening is completely ineffective regardless of $\lambda$}. This indicates another inhibitor for spatial coarsening, which will turn out to be the non-trivial kernel of the curl operator.\\

This puzzling finding made us question, if the implementation was really bug-free, while the explanation to this problem is in fact purely mathematical.

\subsection{An Informal Explanation}\label{Informal}
This subsection explains the underlying mechanics in an intuitive, informal way. The basis for this explanation is the test equation
\begin{equation}\label{TestEquation}
	u'+\lambda u=f~~~~\lambda\geq 0.
\end{equation}
The case  $\lambda = 0$ is the worst case for the convergence of our time parallel scheme assuming some fixed step size. For $\lambda \gg 1$ the smoother converges fast without a coarse grid correction. The semi-discretized eddy-current equation
\begin{equation*}
	\underline{u}'+M_h^{-1}K_h \underline{u} = \underline{f}
\end{equation*}
can be diagonalized by the Eigenvectors of $M_h^{-1}K_h$, which leads with to the test equation with the eigenvalues in place of $\lambda$. This implies that the smoother itself will dampen error components with high eigenvalues. As shown by the local Fourier analysis for differential operators \cite{MR1807961}, high eigenvalues are associated with high spatial frequencies. Dampening those high spatial frequencies smooths the error, which enables spatial coarsening during the coarse grid correction.


For simply connected domains, the eddy-current equation can be viewed through the scope of the Helmholtz decomposition \cite{Helmholtz-Decomposition}
\begin{equation*}
	H(\Omega,\text{curl})= \nabla H^1(\Omega) \oplus \text{curl}~H(\Omega,\text{curl}).
\end{equation*}

By inserting the solenoidal part ($u_s\in\text{curl}~H(\text{curl})$) into the eddy-current equation (\ref{EddyCurrent}) without parameters and by using the identities $\text{curl curl}= \nabla \text{div}-\Delta$, $\text{div curl}=0$ we get the resulting equation
\begin{equation*}
	\frac{\partial u_s}{\partial t} - \Delta u_s = f_s.
\end{equation*} 
Analogously inserting the nodal part ($u_n\in\nabla H^1$) and using $\text{curl}\nabla=0$ results in
\begin{equation}\label{PerspectiveNodal}
	\frac{\partial u_n}{\partial t}=f_n.
\end{equation}
We can conclude from this, that the solenoidal part behaves like a heat equation and should be handled well by our method. In contrast the nodal part is equivalent to the worst case of the test equation \ref{TestEquation}. As a result high frequencies of the gradient part are practically not dampened by the smoother, thus there is no spatial smoothing at all, such that coarsening in space renders the coarse grid correction useless.\\

However equation (\ref{PerspectiveNodal}) can be solved by just integrating the right hand side. Conceptually the error in the nodal component can be calculated by the operator
\begin{equation}\label{InformalCorrection}
	\int_{0}^{s}dt\nabla_x \Delta_x^{-1} \text{div}_x (f-\frac{\partial u}{\partial t}).
\end{equation}
The interpretation of this procedure is as follows.
\begin{itemize}
	\item $f-\frac{\partial u}{\partial t}$: Calculate the residual.
	\item $\text{div}_x$: Eliminate the solenoidal part.
	\item $\nabla_x \Delta_x^{-1}$: Revert $\text{div}_x$ for the nodal part.
	\item $\int_{0}^{s}dt$: Integrate the equation.
\end{itemize}
This can be seen as an expansion of the method \cite{MR1654571} with an additional time component. In this notation the nodal auxiliary space correction by R. Hiptmayr would be denoted by

\begin{equation*}
	\nabla (\text{div}~ \kappa \nabla)^{-1} \text{div} (f-(\kappa I + \text{curl}~\mu^{-1}~\text{curl})u).
\end{equation*}

\subsection{Nodal Auxiliary Space Correction}
With the previous insights in mind we can properly implement the nodal auxiliary correction method. If Nedelec elements of any order are used, an appropriate $H^1(\Omega)$ conforming finite element space can be chosen according to \cite{ZarglmayerDiss}. We define $G$ as the discrete gradient matrix from the $H^1(\Omega)$ FE-space to the Nedelec FE-space and $K_n$ as the nodal Laplace stiffness matrix for the operator $\text{div}~\sigma \nabla$. If we also define $\Sigma_t$ as the operator, which sums up all previous and the current time step
\begin{equation*}
	(\Sigma_t \underline{y})_i:=\sum_{j=1}^{i}y_j,
\end{equation*}
then the correction can be expressed as
\begin{equation}\label{AuxNodalSm}
	\underline{x} \leftarrow \underline{x} + G\Sigma_t K_n^{-1} G^T (\underline{f}-L_{\tau,h}\underline{x}).
\end{equation}
In our notation the spatial operators above are applied to each time step separately, so we can omit the usual tensor products ($I\otimes ...$) for more clarity. Furthermore through a time-multigrid approach the already trivial operator $\Sigma_t$ could be made fully time parallel.

\section{Hybrid Smoother}
The aforementioned auxiliary nodal correction (\ref{AuxNodalSm}) denoted by $A$ can be combined with the standard smoother (\ref{StandardSmoother})  denoted by $S$ to form a hybrid smoother. Like matrices the application is read from right to left, so $SA$ is one application of $A$ followed by an application of $S$. We restrict ourselves to have the same smoother during pre- and post smoothing with reversed order. The following local Fourier analysis indicates, that more than one application of $A$ in the hybrid smoother has no apparent benefits, therefore we restrict ourselves to using $A$ just once.

\subsection{Local Fourier Analysis}

The local Fourier analysis is based on the LFA for the 1D heat equation performed in \cite{MR3521549} and extended with LFA symbols for the 2D curl-curl equation from \cite{MR2407139}. The implementation is done in Mathematica \cite{Mathematica}. The full methodology and more detailed results can be found in \cite[chp. 5]{MasterSchwalsberger}. Additionally the nodal auxiliary correction is implemented. In the LFA we assume exact solutions to the spatial problems (implicit Euler step and Laplace problem), however the numerical experiments in chapter \ref{NumExp} show, that this is not required. Further we assume, that the coarse grid correction is solved exactly.\\

\begin{figure}
	\begin{center}
		\includegraphics[width=0.8 \textwidth]{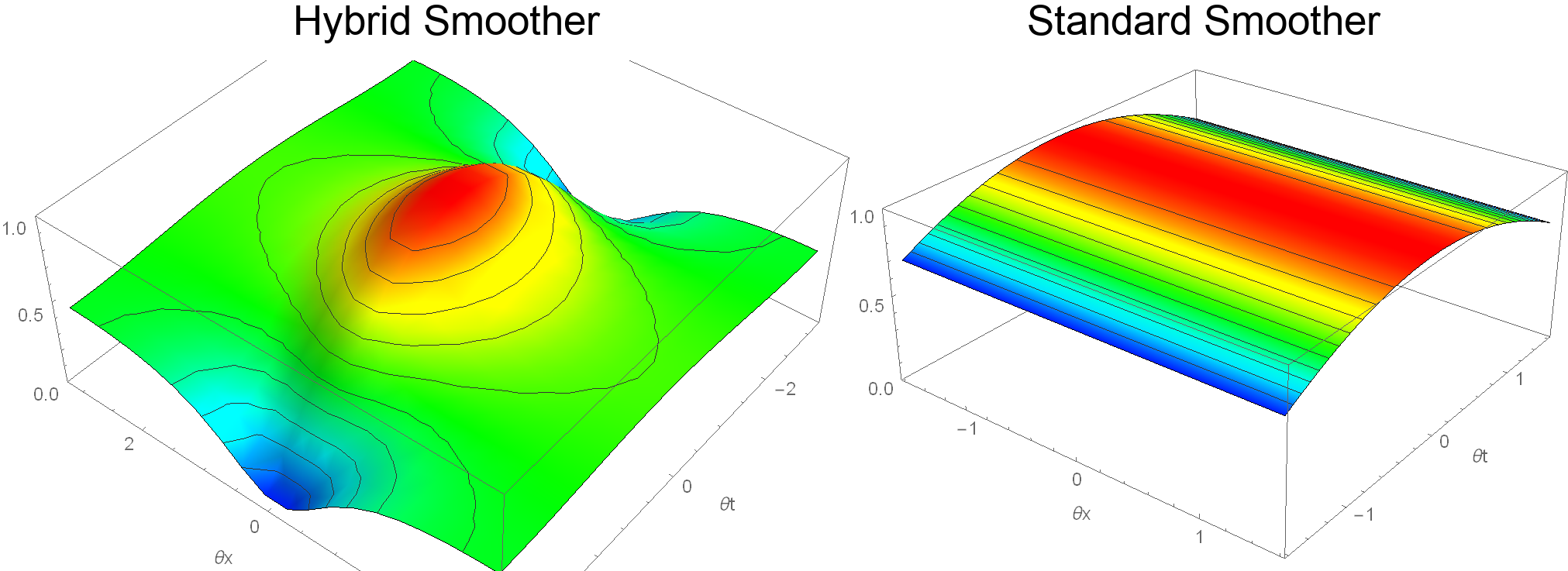}
	\end{center}
	\caption{Comparison between the hybrid smoother $SAS$ and the standard smoother $S$. Spectral radius for each frequency is plotted with $\theta y=0$, $\lambda=1$. Only the hybrid smoother has spatial smoothing properties.}
	\label{LFASmoothers}
\end{figure}

In Figure \ref{LFASmoothers} we observe, how the addition of the auxiliary nodal correction introduces spatial smoothing, whereas the standard smoother features no spatial smoothing at all regardless of $\lambda$ (\ref{CoarseningRule}). This reaffirms our observations in Section \ref{Informal} .\\

\begin{figure}
\begin{center}
		\includegraphics[width=0.8 \textwidth]{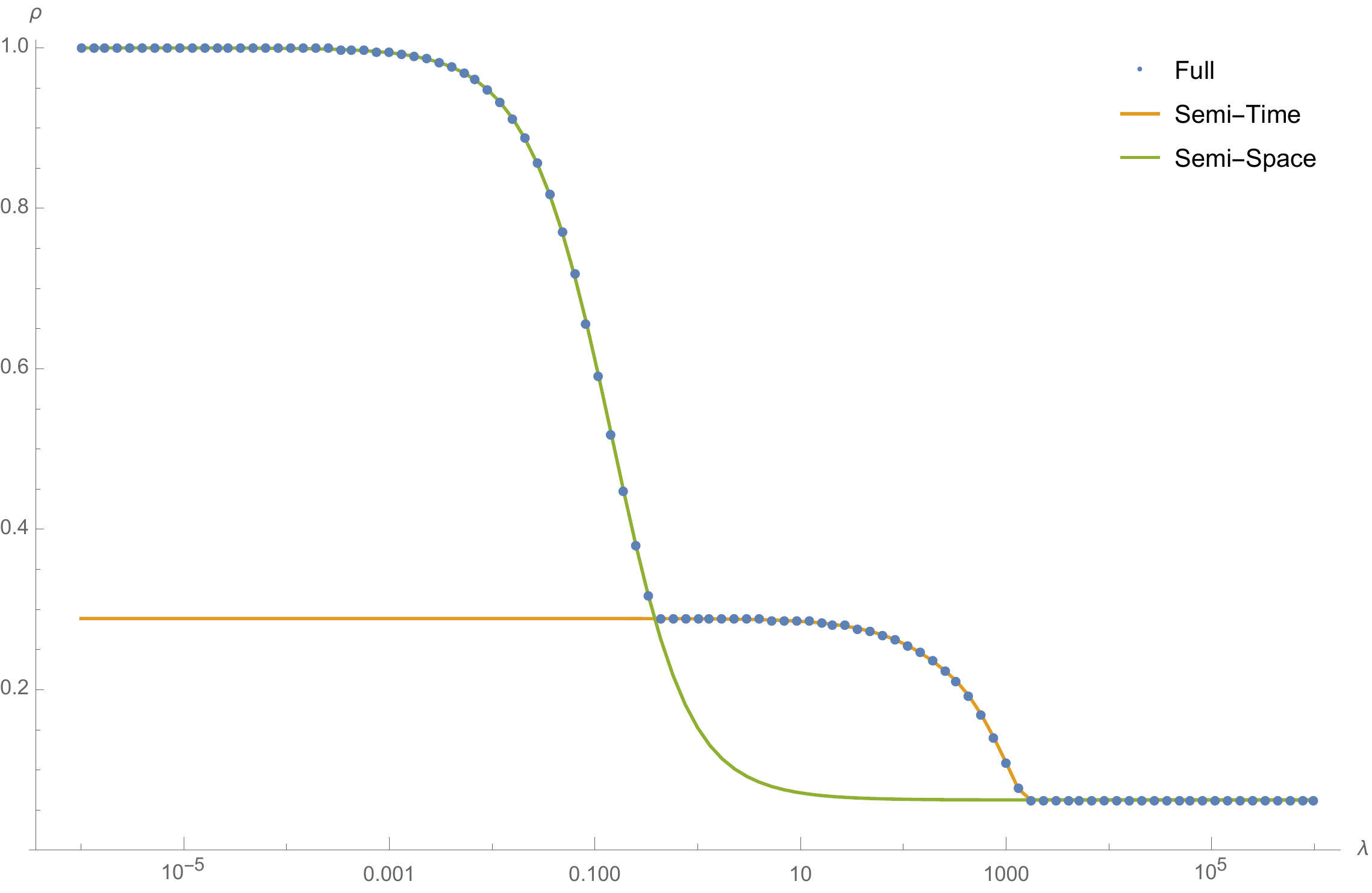}
\end{center}
	\caption{Convergence rates for the two-grid cycle with the $SAS$ smoother as a function of $\lambda$.}
	\label{LFASAS}
\end{figure}

Consequently we measure the maximal spectral radius over all frequencies to get an upper bound for convergence rates. The three investigated methods use either coarsening in space ("Semi-Space"), in time ("Semi-Time"), or both ("Full"). To relate to the coarsening rule, we vary $\lambda$ over the horizontal axis. The results in Figure \ref{LFASAS} confirm, that with the auxiliary smoother in place using a coarser spatial mesh is possible again. Furthermore we can observe, that the "Full" method behaves identical to the "Semi-Time" method for big enough $\lambda$ and deteriorates for lower $\lambda$. Therefore it makes sense to use the coarsening rule (\ref{CoarseningRule}) with $\lambda_{crit}\approx 0.9$ . Not only does the addition of $A$ allow spatial coarsening, but it also improves the convergence rates for "Semi-Time", which would otherwise converge with a worse rate independent of $\lambda$. \\

Different orders like $SSA$ or $ASS$ typically perform slightly worse than $SAS$, therefore we prefer putting $A$ in between $S$. The convergence rates of $SA$ or $AS$ squared are approximately the same as for $SAS$, which means that two V-cycles with $SA$ have the same impact as one V-cycle with $SAS$. This means, that in this setting $SA$ is strictly inferior to $SAS$. On the other end $SSASS$ does not manage to square the convergence rate compared to $SAS$ (0.2 vs. 0.3 for low $\lambda$) except for $\lambda>1000$. In the case of low $\lambda$ we use "Semi-Time", so the coarse grid correction is responsible for half the computational effort and $0.3^{1.337}=0.2$, meaning one V-cycle of $SSASS$ is worth $4/3$ V-cycles of $SAS$. If $SSASS$ was used only on the finest grid, then this would mean approximately $(5/3+1)/2= 4/3$ times the computational effort. This idealized rough estimate could set $SSASS$ on par with $SAS$ in certain cases, maybe $SSAS$ could also be a good choice. In conclusion we will focus on $SAS$ as a solid choice.

\section{Numerical Experiments}\label{NumExp}

Our practical implementation is based on the implementation \cite{MR3521549} and uses MFEM \cite{mfem-library} and HYPRE \cite{hypre-library} to deal with the arising spatial problems. The implementation is parallel in time, while everything except the auxiliary method is parallel in space, which was only prevented by time constraints. Therefore it is close to being parallel in space and time.\\


The modification of the existing multigrid framework to support a different spatial problem was seamless due to the non-intrusive design of the method. Specifically only the application of spatial prolongation/restriction, the discrete gradient matrix and functionality for $A_\tau,M_h,K_n$ needs to be provided through an interface. For example swapping out the eddy-current equation for a heat equation is a matter of minutes and matrix-free methods could be incorporated as well. The multigrid method itself fully relies on the interface for applying and solving spatial operators. The only hard-coded part is the used vector class, which could also be replaced by a template-based approach. There are other methods which aim for even less intrusiveness [REF reduced integration], but we are confident, that for many preexisting applications the required interface could be implemented with ease.\\

\subsection{Determining Convergence Rates and the Parameter $\lambda_{crit}$}

As we are not aware of any available local Fourier analysis for the 3D curl-curl equation, we recreate Figure \ref{LFASAS} by using our implementation of the multigrid method with a $SAS$ smoother. In addition we want to showcase how a fitting $\lambda_{crit}$ can be determined with nothing but a working implementation. To determine convergence rates the power iteration method was applied to a test case. This means, that we apply up to 200 iterations of the method to a random initial guess (values between 0 and 1) with a $0$ right hand side. During this process we measure the reduction of the $\ell^2$ norm of the vector, which converges to the spectral radius of the method. The maximal occurring error reduction factor is considered as convergence rate, because these factors are almost monotonically increasing. In turn means, that our method has no worse convergence during a pre-asymptotic phase. Because we use no rescaling, the iteration is stopped, if the total error reduction gets below $10^{-100}$ to avoid a double underflow.\\

We compare the time multigrid method without coarsening in space ("Semi-time") with a variant, which uses just two levels of the spatial mesh and uses full coarsening on the highest level ("Full") as in Table \ref{FullMethod}. The reason is that the local Fourier analysis assumes the coarse grid correction to be solved exactly (two-grid cycle), so we mimic this by not using further spatial coarsening within the correction. For "Full" we want to use the spatial coarsening right on the highest level, because potentially bad convergence rates on lower levels are obscured on the highest level, where the convergence rates are measured.\\

\begin{table}
\begin{center}
		\begin{tabular}{|c|c|c|c|c|c|c|c|c|c|c|}
		\hline 
		Space/Time level & 9 & 8 & 7 & 6 & 5 & 4 & 3 & 2 & 1 & 0 \\ 
		\hline 
		2 & x &  &  &  &  &  &  &  &  &  \\ 
		\hline 
		1 &  & x & x & x & x & x & x & x & x & x \\ 
		\hline 
		0 &  &  &  &  &  &  &  &  &  &  \\
		\hline
	\end{tabular} 
\end{center}
	\caption{Coarsening strategy example for the "Full" method test case.}
	\label{FullMethod}
\end{table}

The basic mesh consists of two tetrahedra, which are uniformly refined several times. The problem features homogeneous Neumann boundary conditions. We compare the results for $2,3,4,5,6$ uniform refinements, so we can observe convergent behavior with an increasingly fine mesh. The spatial degrees of freedom range from 230 to 630240, with 512 time steps. The material parameters $\sigma, \mu$ are set to 1 and $\lambda$ is varied by using different values for $\tau$.\\

For solving the spatial problems $A_{\tau,h},K_n$ we rely on 3 iterations of the preconditioned conjugate gradient method. The preconditioners are algebraic multigrid methods from HYPRE \cite{hypre-library}. BoomerAMG was used for the Laplace problem and AMS for $A_{\tau,h}$. Quite remarkably, turning off the gradient-smoothing in AMS has no impact on convergence, because the gradient part is handled by the auxiliary correction. At the same time this cuts down total computation times by roughly $30\%$.\\

Our main goal is to confirm, that the method behaves similar to the local Fourier analysis for the 2D curl-curl equation. Furthermore we want to determine the value of $\lambda_{crit}$ in this setting by observing, where the two methods "Full" and "Semi-time" start to differ.\\

\begin{figure}
	\begin{center}
		\includegraphics[width=0.8 \textwidth]{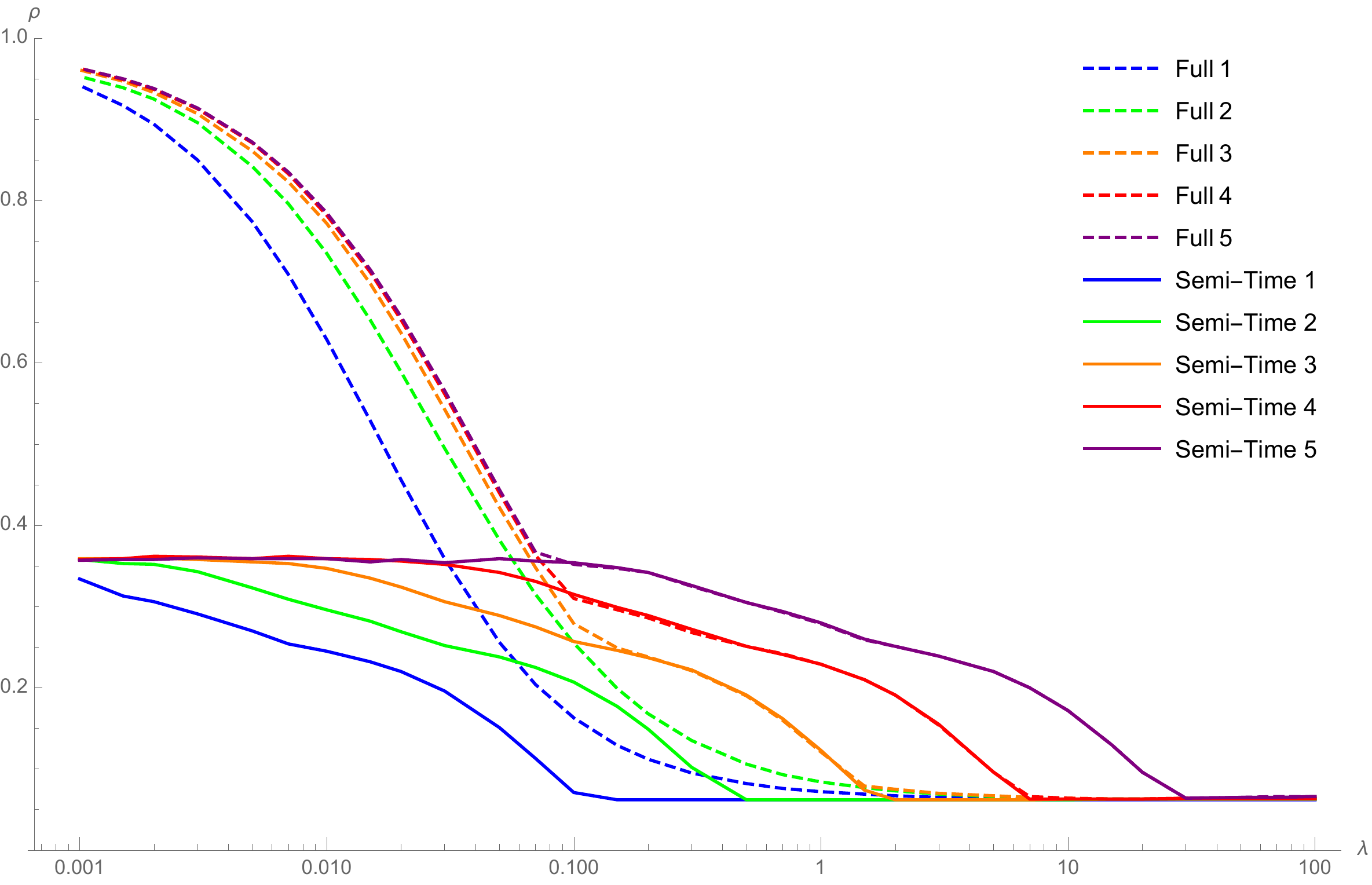}
	\end{center}
	\caption{Convergence rates for the two-grid cycle with $SAS$ smoother as a function of $\lambda$. Comparision between "Full" and "Semi-time" for different numbers of uniform refinements.}
	\label{ExperimentLC}
\end{figure}

\begin{figure}
	\begin{center}
		\includegraphics[width=0.8 \textwidth]{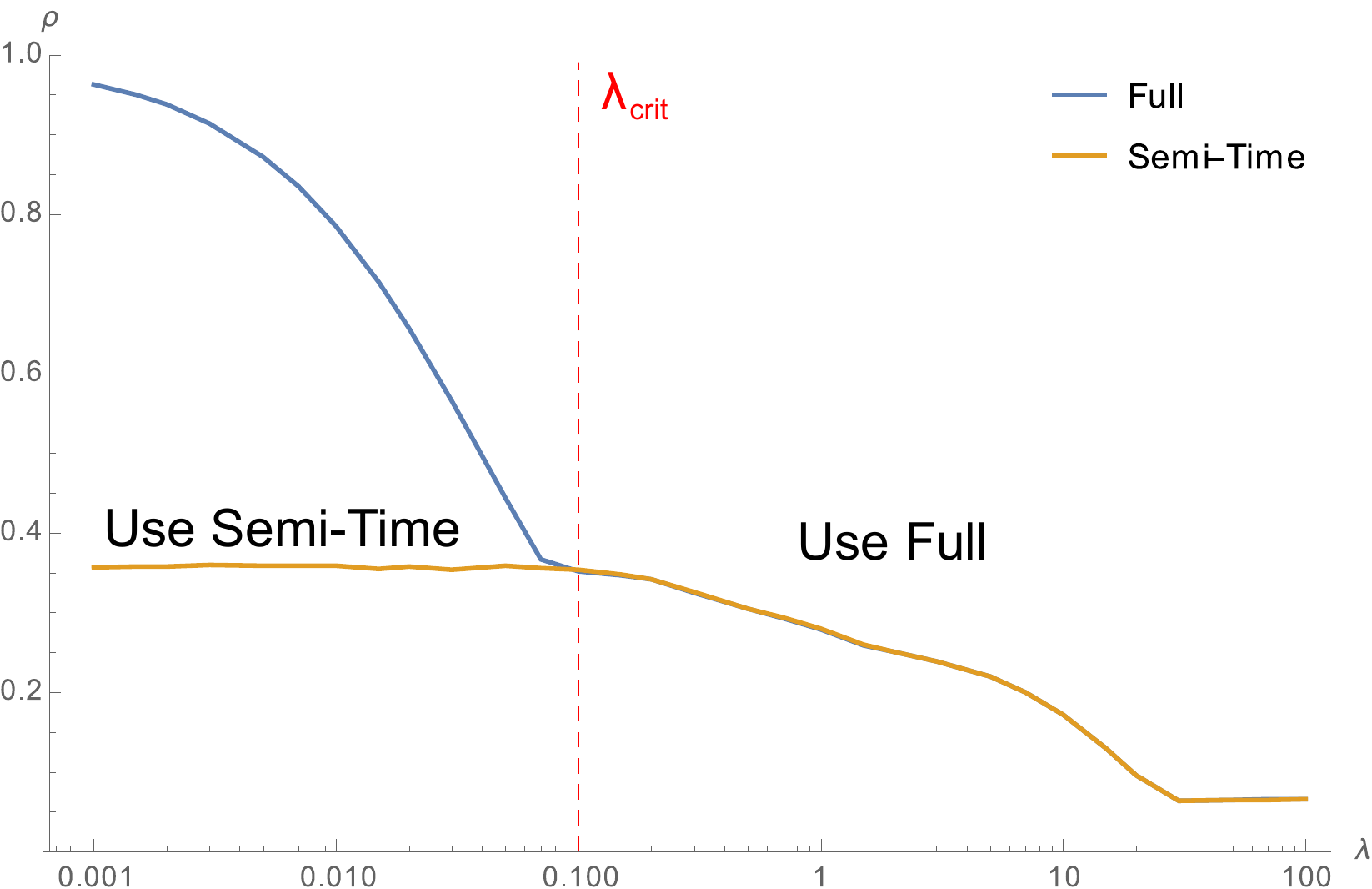}
	\end{center}
	\caption{Convergence rates for the two-grid cycle with $SAS$ smoother as a function of $\lambda$. Only the most refined cases with 630240 DoF in space. $\lambda_{crit}$ determined as $0.1$.}
	\label{ExperimentLCR5}
\end{figure}

In Figure \ref{ExperimentLC} we can observe that the important parts of the curves converge with an increasing number of refinements. The only non-convergent feature is the transition from the high $\lambda$ region with a convergence rate of $2^{-4}$ to the plateau of the "Semi-time" method with a convergence rate of $0.36$. 
 This means, that our method performs \textit{better than expected} in this regard. On a side note the smoother on its own shifts to the right in the same fashion and is a suitable solver for sufficiently high $\lambda$ and small problems. In the case of the heat equation the smoother never converges on its own and none of the shifting happens.\\

Figure \ref{ExperimentLCR5} omits all cases except the most refined one for the sake of a less cluttered plot.  The point where the "Full" method departs from the "Semi-time" method can be located at 0.1. When the diagonal line is extrapolated, it even appears to be around 0.075. This means that a choice of $\lambda_{crit}=0.1$ will safely guarantee the optimal coarsening strategy to be selected. This holds for the case of the 3D eddy-current equation, lowest order Nedelec elements, implicit Euler time stepping, $SAS$ smoother, $h=$ longest edge.\\

\subsection{Coarsening Rule for Locally Different Parameters}

In practical applications vastly different material parameters are very common. For example the ratio $\frac{\mu^{-1}}{\sigma}$ can differ by a factor of $\approx 30$ between iron and copper, while between iron and weak conductors this gap can be way bigger. The limit case of perfect insulators ($\sigma = 0$) would even cause an infinitely large local degree of anisotropy $\lambda$. This motivates the question, if the local nature of the coarsening rule (\ref{CoarseningRule}) can indeed handle different materials. On a side note we are not concerned, if the preconditioner can handle jumping coefficients, because our space-time parallel method can use any suitable method to solve the spatial problems.\\

We assign two different material parameters to the two original tetrahedra of the test mesh. After uniform refinements two regions with two distinct degrees of anisotropy $\lambda_1<\lambda_2$ are given. As our method mostly benefits from increasing $\lambda$, it is sufficient to only deal with the smallest $\lambda$ as reflected in the coarsening rule (\ref{CoarseningRule}).\\

In contrast the method in \cite{MR1335894} chooses to use semi coarsening in time if $\lambda > \lambda_{crit}$ and semi coarsening in space if $\lambda < \lambda_{crit}$. If this rule is violated both semi coarsening in time (when $\lambda < 2^{-4}\lambda_{crit}$) and semi coarsening in space (when $\lambda > 2^{4}\lambda_{crit}$) yield bad convergence rates ($>0.8$). Therefore we are concerned, that in the case of two materials with $\lambda_1\leq 2^{-5},\lambda_2\geq 2^{3}$ neither coarsening approach would yield useful convergence rates. Vastly different $\lambda_i$ can also arise during adaptive refinements, where the size of the elements can differ significantly.\\

For this "Two $\lambda$" test case we use mostly the same setup as for the previous test case, but we only investigate the "Full" method, only use 4 uniform refinements, have a maximum of 100 iterations and vary $\mu^{-1} \geq 1$ on one of the two subdomains. This has no effect on the minimal $\lambda$, as it only increases $\lambda_2$. The minimal $\lambda$ is again manipulated through changing $\tau$.\\

\begin{table}
	\begin{center}
	\begin{tabular}{|c|c|c|c|c||c|c|c|c|}
		\hline 
		$\mu^{-1}_2\setminus\lambda$& $2^0$ & $2^1$ & $2^2$ & $2^3$ & $2^4$ & $2^5$ & $2^6$ & 0.01 \\ 
		\hline 
		1& 0.121 & 0.19 & 0.23 & 0.252 & 0.372 & 0.533 & 0.688 & 0.77 \\ 
		\hline 
		$10^1$& 0.083 & 0.159 & 0.213 & 0.242 & 0.34 & 0.498 & 0.662 & 0.753 \\ 
		\hline 
		$10^2$& 0.083 & 0.158 & 0.212 & 0.242 & 0.34 & 0.5 & 0.664 & 0.755 \\ 
		\hline 
		$10^3$& 0.083 & 0.158 & 0.212 & 0.242 & 0.34 & 0.5 & 0.665 & 0.756 \\ 
		\hline 
		$10^4$& 0.083 & 0.158 & 0.212 & 0.243 & 0.34 & 0.501 & 0.665 & 0.756 \\ 
		\hline 
		$10^5$& 0.083 & 0.157 & 0.212 & 0.243 & 0.34 & 0.501 & 0.665 & 0.756 \\ 
		\hline 
		$10^6$& 0.083 & 0.157 & 0.213 & 0.242 & 0.34 & 0.501 & 0.665 & 0.756 \\ 
		\hline 
	\end{tabular} 
	\end{center}
\caption{Maximum error reduction factor for the "Two $\lambda$" test case using the "Full" method. 10744 DoF in space. After an initial boost to convergence, the rates remain stable for increasing $\mu_2^{-1}$. Again $\lambda_{crit}$ appears to be located between 0.06125 and 0.125.}
\label{TestCase2}
\end{table}

We can observe in Table \ref{TestCase2}, that only the minimal $\lambda$ has a notable impact on convergence, while increasing the bigger $\lambda_2$ can only improve the convergence rate. With this and previous results \cite{MasterSchwalsberger} our method is one step closer to real applications with varying materials.

\section{Parallel Performance}
To investigate the parallel performance of the method, we perform a weak scaling test with respect to time. In contrast to other weak scaling tests we expand the time domain instead of refining it. This is done to keep the degree of anisotropy constant at $\lambda=0.08$, which is allows for instant spatial coarsening. For comparison we also run a test series without spatial coarsening.

\begin{itemize}
	\item Spatial domain $(0,1)^3$, 7768 DoF
	\item $2^k$ time-parallel processors
	\item Time domain $32*2^k$ time steps with size $0.016$
	\item $\Rightarrow$ Constant $\lambda = 0.512$
	\item $\lambda_{crit} = 0.15$
	\item Material parameters = 1
	\item Residual reduction tolerance $10^{-10}$
	\item SAS smoother
\end{itemize}

The "Full" method always finished in 12 iterations, while the "Semi-Time" method needed 14 iterations. Without some of the idealizations assumed in the analysis it is very well possible, that the "Full" method has a marginally better convergence rate than the "Semi-Time" method. The corresponding run-times can be observed in Table \ref{WeakScaling}. The Adjusted Ratio is the ratio of the runtimes multiplied by $14/12$ to adjust for the iterations.

\begin{table}[h]
	\small
	\begin{center}
		\begin{tabular}{|c|c|c|c|c|c|c|c|c|c|c|}
			\hline 
			\#Processors & 1 & 2 & 4 & 8 & 16 & 32 & 64 & 128 & 256 & 512 \\ 
			\hline 
			Full & 35.3 & 35.3 & 35.4 & 39.3 & 49.8 & 49.7 & 50.2 & 50.2 & 50.4 & 52.1 \\ 
			\hline 
			Semi-Time & 74.6 & 75.8 & 77.4 & 87.6 & 112.3 & 114.1 & 115.6 & 117.6 & 118.9 & 120.8 \\
			\hline 
			Adjusted Ratio & 0.552 & 0.543 & 0.534 & 0.523 & 0.517 & 0.508 & 0.507 & 0.498 & 0.495 & 0.503 \\
			\hline
		\end{tabular} 
	\end{center}
	\caption{Weak scaling results in seconds for the "Full" and the "Semi-Time" method. Adjusted ratio is Full/Semi-Time*14/12}
	\label{WeakScaling}
\end{table}

We can clearly observe how the "Full" method scales better with barely any increase in computation times in the upper range. The sudden increase around the 16 cores mark can be most likely attributed to the architecture of the cluster with a node size of 16 cores.\\

It would have been interesting to run the scaling tests on even bigger clusters, where possibly a more severe discrepancy between the two methods would surface.

\section{Conclusion}
This paper demonstrates the difficulties in solving the eddy-current equation with a space-time multigrid method using also spatial coarsening. The non-trivial kernel of the curl-operator can be identified as the cause of these problems, and a nodal auxiliary space correction with time integration is devised to restore good convergence rates. This procedure is analogous to the method in \cite{MR1654571}, but with an additional time component.\\

The local Fourier analysis of the original and the modified method confirms the effectiveness of the auxiliary correction, and in fact lead the authors of this paper to devise the auxiliary correction. From the gained insights, we expect similar problems to occur whenever the spatial operator features a non-trivial kernel. Furthermore, we expect other space-time multigrid methods to experience similar problems and to profit from an implementation of the described universal correction (\ref{InformalCorrection}). Even multigrid methods without spatial coarsening could benefit from this modification, because it also improves the convergence rates of our pure time-multigrid method.\\

We proceed to determine convergence rates and $\lambda_{crit}=0.1$ with numerical tests. This result holds for the 3D eddy-current equation and one specific configuration of the method. The coarsening rule passes one basic test for robustness, which involves solving test problems with vastly different local degrees of anisotropy $\lambda_i$.\\

We finally demonstrate with the help of weak scaling tests, how spatial coarsening can improve parallel performance.\\

In conclusion the explored "nodal auxiliary space correction" is a highly useful tool to enhance space-time multigrid methods for the eddy-current equation.

\section*{Acknowledgment}
The research was funded by the Austrian Science Fund (FWF) under the grant W1214-N15, project DK8. 
We further want to thank the RICAM institute for enabling computations on the RADON1 cluster and the people associated with the NUMA institute for helpful comments and insights. We want to especially thank Ulrich Langer for fruitful discussions. Thank also goes to Bianca Klammer for additional proofreading.




%
%
\newpage

\bibliographystyle{acm}
\bibliography{References}

\end{document}